# Significant Radiation Enhancement in Photoconductive Terahertz Emitters by Incorporating Plasmonic Contact Electrodes

C. W. Berry [(1)], M. R. Hashemi, M. Unlu, and M. Jarrahi [(2)]
Electrical Engineering and Computer Science Department, University of Michigan
Ann Arbor, MI 48109, United States

Email addresses: (1) berryc@umich.edu, (2) mjarrahi@umich.edu

**ABSTRACT** Even though the terahertz spectrum is well suited for chemical identification, material characterization, biological sensing and medical imaging, practical development of these applications has been hindered by the attributes of terahertz sources, namely low output power and poor efficiency. Here, we demonstrate that use of plasmonic contact electrodes can significantly enhance the optical-to-terahertz conversion efficiency in a photoconductive terahertz emitter. The use of plasmonic contact electrodes offers nanoscale carrier transport path lengths for the photocarriers, enabling efficient collection of the majority of carriers in a sub-picosecond time-scale. It also allows increasing photoconductor active area without a considerable increase in the capacitive loading to the antenna, boosting the maximum terahertz radiation power by preventing the carrier screening effect and thermal breakdown at high optical pump powers. We experimentally demonstrated 50 times higher terahertz radiation power from a plasmonic photoconductive emitter in comparison with a similar photoconductive emitter with non-plasmonic contact electrodes.

There has been a great deal of effort for extending the high frequency limit of RF sources and the low frequency limit of optical emitters as well as combining RF and optical techniques to realize high performance terahertz radiation sources [1]. On the RF side, IMPATT diodes, Gunn diodes, resonance tunneling diodes [2, 3], and chains of frequency multipliers [4, 5] have demonstrated very promising compact terahertz sources. However, this category of sources has limited bandwidth, poor power efficiency and low output power levels. On the other hand, electron beam devices such as backward wave oscillators [6] and travelling wave tube regenerative amplifiers [7] can produce reasonable power levels, but their operation has not been demonstrated above 1.5 THz [8]. Additionally, the bulky nature of backward wave oscillators and their requirement for high magnetic fields and vacuum limits their use in various operational settings. On the optical side, quantum-cascade lasers have been under an extensive investigation during the past decade [9], and significant progress has been made pushing their operation frequency to 1 THz [10] and their operation temperature to ~200 K for a 3.22 THz lasing frequency [11]. Additionally, optical down-conversion to terahertz frequencies based on nonlinear optical effects has been extensively used for generating high power terahertz waves. Optical down-conversion to terahertz frequencies in bulk nonlinear materials is inherently inefficient due to the optical/terahertz phase mismatch limiting the efficient field interaction length [12]. Guided wave nonlinear media [13-15], quasi-phase-matching in periodically poled media [16-21], and the use of tilted wave-front pump waves [22] have been employed to offer a better phase-matching control resulting in longer field interaction lengths. However, because of material absorption, the active length in which THz waves are generated is limited to centimeter ranges. Because of the field interaction length limitations, the use of high-power optical pumps has been necessary to provide ultra-high peak powers for generating meaningful terahertz powers using nonlinear optical techniques. An additional inherent limitation of nonlinear optical techniques for generating terahertz waves stems from the conservation of the number of photons in a nonlinear optical process, namely the Manley–Rowe rule [23]. In other words, the maximum power efficiency of terahertz sources based on nonlinear optical phenomena is limited to the ratio between the energies of the generated terahertz photon and the pump optical photon.

On the other hand, optical-to-terahertz conversion through photoconduction has demonstrated very promising performance [24-30] and has been the most commonly used technique for generating terahertz waves since the pioneering demonstration of picosecond photoconducting Hertzian dipoles in 1984 [31]. One of the main advantages of photoconductive terahertz emitters compared to the terahertz emitters based on nonlinear optical phenomena is that their power efficiency is not restricted by the Manley-Rowe limit. This is because one electron-hole pair can be generated for each absorbing photon, which can emit several terahertz photons upon reaching the terahertz antenna in a photoconductive emitter. In other words, the optical-to-terahertz conversion efficiency of photoconductive terahertz emitters can reach 100%, orders of magnitude higher than the Manley-Rowe limit. Although the optical-to-terahertz conversion efficiency of photoconductive emitters can theoretically exceed 100%, the low quantum efficiency of conventional ultrafast photoconductors imposes substantially lower conversion efficiencies. To address the power efficiency limitation of conventional photoconductive terahertz emitters, we have proposed a novel photoconductive terahertz emitter concept which incorporates a plasmonic contact electrode configuration to offer high quantum-efficiency and ultrafast operation simultaneously. By using nano-scale plasmonic contact electrodes we significantly reduce the average photogenerated carrier transport path to photoconductor contact electrodes compared to conventional photoconductors. By incorporating plasmonic contact electrodes, we demonstrate enhancing the optical-to-terahertz power conversion efficiency of a conventional photoconductive terahertz emitter by a factor of 50.

**Results**
To demonstrate the potential of plasmonic electrodes for terahertz power enhancement, we chose to fabricate a proof-of-concept photoconductive emitter and characterized its performance enhancement by integrating plasmonic gratings as a part of photoconductor contact electrodes. Figure 1a shows the schematic diagram and operation concept of the implemented photoconductive emitter in the absence of plasmonic gratings (conventional scheme). The photoconductive emitter consists of an ultrafast photoconductor with 20 μm gap between anode and cathode contacts, connected to a 60 μm long bowtie antenna with maximum and minimum widths of 100 μm and 30 μm, respectively. Low-temperature-grown GaAs (LT-GaAs) is used for the photo-absorbing substrate to achieve an ultrafast photoconductor response. When a sub-picosecond optical pump at ~ 800 nm wavelength range is incident on the ultrafast photoconductor, electron-hole pairs are generated which induce a photocurrent under an applied bias electric field. The induced photocurrent, which follows the envelope of the optical pump, drives the terahertz antenna connected to the photoconductor contact electrodes, generating terahertz radiation. The bandwidth of terahertz radiation is determined by the optical pulse, antenna characteristics, and the semiconductor carrier lifetime. Since electrons have significantly higher mobilities compared to holes and due to the nonlinear increase in the bias electric field near contact electrodes, the optical pump is focused onto the photoconductive gap asymmetrically close to the anode contact to maximize terahertz radiation [27]. Similar to any conventional photoconductive terahertz emitter, the quantum efficiency of the described photoconductive emitter is limited by the relatively long carrier transport path lengths to photoconductor contact electrodes. This is because the drift velocity of electrons cannot exceed $2\times10^7$ cm/s [32, 33] and even if the pump is focused down to a diffraction-limited spot size next to the anode contact, a very small portion of the photo-generated electrons can reach the anode in a sub-picosecond timescale. The remaining majority of photo-generated carriers recombine in the substrate before reaching the contact electrodes and without efficient contribution to terahertz generation. Figure 1b shows the schematic diagram of the implemented photoconductive emitter that incorporates 20 μm long plasmonic contact electrode gratings designed for enhancing the quantum efficiency of the conventional photoconductive emitter (Fig. 1a). The grating geometry (200 nm pitch, 100 nm spacing, and 50 nm height Au grating) is designed to excite surface plasmon waves along the periodic metallic grating interface upon incidence of a TM-polarized optical pump [34, 35]. Excitation of surface plasmon waves allows transmission of more than 70% of the optical pump through the nano-scale grating into the photo-absorbing substrate. It also enhances the intensity of the optical pump in very close proximity to Au electrodes significantly. As a result, the average photo-generated electron transport path length to the anode electrode is significantly reduced in comparison with the conventional photoconductive emitter (Fig. 1a).

Using a finite element solver (COMSOL), we have analyzed the interaction of an incident optical pump ($\lambda$ = 800 nm) with the conventional and plasmonic photoconductors. For maximum optical power enhancement near photoconductor contact electrodes, we have aligned the optical pump field along the *x*-axis and *y*-axis for the conventional and plasmonic photoconductors, respectively. The optical absorption in the semiconductor substrate for the conventional (*xz* cross section) and plasmonic (*yz* cross section) photoconductors is shown in Fig. 2a. For the conventional photoconductor, the metal contact shadows the substrate from the incident optical pump, allowing for almost all of the optical absorption and photocarrier generation in the gap between the anode and cathode. In the case of the plasmonic photoconductor, the optical pump is transmitted through the nanoscale metallic grating through the coupling with surface plasmons. Since the excited surface plasmon waves exist at the dielectric-metal interface, the highest optical absorption and photocarrier generation occurs in direct proximity to the metal contacts. Since the saturation velocity of electrons in GaAs is approximately $1\times10^7$ cm/s [32, 33], only photocarriers generated within 100 nm of the electrodes will be able to travel to the contacts within 1 ps. Simulation results show that while 38% of the absorbed photons generate photocarriers within 100 nm from the plasmonic photoconductor anode contact, just 4% of the absorbed photons generate photocarriers within 100 nm from the conventional photoconductor anode contact for a diffraction-limited pump focus.

Using COMSOL, we have analyzed the bias electric field that drifts photogenerated carriers toward the photoconductor contact electrodes (Fig. 2b). For this analysis, the bias voltage is set such that the maximum induced electric field remains below $10^5$ V/cm (1/4$^{th}$ of the GaAs breakdown electric field). In the case of the conventional photoconductor, we see elliptical electric field lines beneath the contact electrode with the highest electric field near the corner of the electrodes. In the case of the plasmonic photoconductor, the electric field lines are illustrated at the *yz* cross section 1 µm away from the grating tip. Although relatively lower electric fields are induced in the case of the plasmonic photoconductor compared with the conventional photoconductor, the electric field levels are maintained above $10^3$ V/cm within 100 nm from photoconductor anode contact electrodes. Therefore, the drift velocity of electrons generated within 100 nm from the anode contact is approximately $1\times10^7$ cm/s (saturation velocity) for both the conventional and plasmonic photoconductors [32, 33]. Similar electron drift velocity levels for the conventional and plasmonic photoconductors combined with ~1 order of magnitude higher photocarrier generation within 100 nm of contact electrodes in the case of the plasmonic photoconductor results in ~1 order of magnitude more photocurrent driving the antenna of the plasmonic photoconductive emitter. Therefore, the terahertz power radiated from the plasmonic photoconductive emitter is expected to be ~2 orders of magnitude higher than the conventional photoconductive emitter. Moreover, simulation results show that the electric field levels are maintained above $10^3$ V/cm within 100 nm from contact electrodes along a 20 µm long plasmonic contact electrode, indicating that the superior performance of the plasmonic photoconductive emitter can be maintained when using relatively large device active areas.

Figure 3a shows the microscope and SEM images of the fabricated photoconductive emitter prototypes. Fabrication began with the patterning of the nanoscale gratings using electron beam lithography, followed by Ti/Au (5/45 nm) deposition and liftoff. A 150 nm thick SiO$_2$ passivation layer was then deposited using PECVD. Using a plasma etcher, contact vias were then opened and the antenna and metal contacts were formed using optical lithography followed by Ti/Au (10/400 nm) deposition and liftoff. The fabricated devices were then mounted on a silicon lens and placed on an optical rotation mount, for optical pump polarization adjustments. To evaluate the performance of the photoconductive terahertz emitter with and without the plasmonic gratings, the output power of each device was measured using a pyroelectric detector (Spectrum Detector, Inc. SPI-A-65 THz). The incident optical pump from a Ti:sapphire laser with a central wavelength of 800 nm, 76 MHz repetition rate, and 200 fs pulse width was tightly focused onto each device and positioned to maximize the measured radiation. By adjusting the device rotation mount, the electric field of the optical pump was aligned along the *x*-axis and *y*-axis for the conventional and plasmonic prototypes, respectively (as illustrated in Fig. 1).

The measured output power of the two prototype devices at a 40 V bias voltage and under various optical pump powers is presented in Fig. 3b. A radiation power enhancement of more than 33 was observed from the plasmonic photoconductive emitter in the 0 - 25 mW optical pump power range. This significant radiation power enhancement is due to the higher photocurrent levels generated when employing plasmonic contact electrodes (Fig. 3b inset), which have a quadratic relation with the radiation power. To further examine the impact of photocurrent increase on the radiation power enhancement, we compared the radiation power of the two photoconductor devices as a function of their photocurrent under various bias voltages (10 - 40 V) and optical pump powers (5 - 25 mW). The results are presented in Fig. 3c in a logarithmic scale. The data points are all curve-fitted to the same line with a slope of 2, confirming the quadratic dependence of the radiation power on the induced photocurrent and the fact that all other operational conditions (including antenna specifications) are the same for the conventional and plasmonic photoconductive emitter prototypes. By dividing the output power of the plasmonic photoconductive emitter by the output power of the conventional photoconductive emitter, we define a power enhancement factor. Figure 3d shows the power enhancement factor under various optical pump powers and bias voltages. At low optical pump power levels and a bias voltage of 30 V, output power enhancement factors up to 50 are observed, the same order of magnitude predicted by the theoretical predictions. The enhancement factor decreases slightly at higher optical pump power levels and higher bias voltages. This can be explained by the carrier screening effect, which should affect the plasmonic photoconductor more than the conventional photoconductor, since it is generating more photocurrent and separating a larger number of electron-hole pairs. Finally, the maximum radiated power from each photoconductive terahertz emitter was measured at an optical pump power of 100 mW, up to the point that the devices burned, as shown in Fig. 3e. At maximum, the plasmonic photoconductive emitter produced an average power of 250 μW, compared to the 12 μW of the conventional photoconductive emitter.

The time-domain and frequency domain radiation of the plasmonic photoconductive emitter was measured using a terahertz time-domain spectroscopy setup with electro-optic detection [36]. The time-domain radiation power is shown in Fig. 4a, indicating a radiation pulse with a full width at half maximum of 660 fs in response to a 200 fs optical pump pulse from the mode-locked Ti:sapphire laser. The radiated power from the plasmonic terahertz source was detected up to 1.5 THz, after which the noise of the system limited detection (Fig. 4b). The radiation peaks observed at 0.5 THz, 0.25 THz, and 0.125 THz are associated with the resonant nature of the employed bow-tie antenna as the radiating element of the photoconductive emitter prototypes.

**Discussion**

The ability to excite surface plasmon waves has enabled many unique opportunities for routing and manipulating electromagnetic waves [37]. It has enabled strong light concentration in the near-field paving the way for higher resolution imaging and spectroscopy [38, 39], deep electromagnetic focusing and beam shaping [40, 41], higher efficiency photovoltaics [42, 43], photodetectors [44, 45], modulators [46, 47], and photoconductors [30]. On the basis of this capability, we have proposed and experimentally demonstrated two orders of magnitude radiation power enhancement by incorporating plasmonic contact electrodes in a photoconductive terahertz emitter. Plasmonic electrodes are designed to significantly reduce the average carrier transport distance to photoconductor contact electrodes over relatively large device active areas and without a considerable increase in the capacitive loading to the terahertz radiating antenna. This enables boosting the maximum terahertz radiation power [48-52] by mitigating the carrier screening effect [53], semiconductor bleaching [54], and thermal breakdown [55] at high optical pump powers.

It should be noted that the focus of this study has been the demonstration of the impact of plasmonic electrodes in enhancing the induced photocurrent in ultrafast photoconductors and the radiated terahertz power from photoconductive terahertz emitters. Thus, the choice of the terahertz radiating antenna and bias feed in our study has been arbitrary, and the enhancement concept can be similarly applied to enhance the radiation power from photoconductive terahertz emitters with a variety of terahertz antennas with and without interdigitated contact electrodes in both pulsed and continuous-wave operation. In this regard, the output power of our prototype devices can be further enhanced through use of resonance cavities [26, 29] and antennas with higher radiation resistance and bandwidth [56, 57]. Moreover, the use of high aspect ratio plasmonic contact electrodes embedded inside the photo-absorbing semiconductor [58-60] allows a larger number of carriers generated in close proximity with photoconductor contact electrodes and, thus, enables further terahertz radiation enhancement. In this regard, extending the plasmonic electrode height to dimensions comparable with the optical pump absorption depth would eliminate the need for using short carrier lifetime semiconductors which have lower carrier mobilities and thermal conductivities [55] compared to high quality crystalline semiconductors.

## Methods

In performing terahertz radiation power measurements, the beam from a mode-locked Ti:sapphire laser (800 nm central wavelength, 76 MHz repetition rate, 200 fs pulse width, 5 Hz amplitude modulation) is tightly focused onto the active area of the device under test. The device, mounted on an optical rotation mount and held by an XYZ translation stage, is positioned to maximize the radiated power. For optimal performance of the conventional terahertz emitter, the incident optical electric field is oriented to span the anode-cathode gap. The metal gratings of the plasmonic terahertz emitter are likewise oriented to run perpendicular to the incident electric field, for best performance. A parametric analyzer is used to simultaneously apply bias voltages up to 40 V and measure the electrical current drawn. Emitted radiation is measured using a pyroelectric detector (Spectrum Detector, Inc. SPI-A-65 THz) and converted to terahertz power using the manufacturer's responsivity values.

Emitted terahertz power is monitored in time-domain and frequency domain through a terahertz time-domain spectroscopy setup. The setup uses the same Ti:sapphire laser used in the power measurements with the pump path amplitude modulated at 2 KHz. The pump beam is focused onto the device under test and emitted radiation is collimated and subsequently focused using two polyethylene spherical lenses in the ambient atmosphere. Before the focal point of the second polyethylene lens, the terahertz radiation is combined with the optical pump beam using an ITO coated glass filter. Both paths are then focused onto a 1 mm thick, <110> ZnTe crystal. The optical beam is then converted into circular polarization, separated by a Wollaston prism, and measured using two balanced detectors connected to a lock-in amplifier.

## Acknowledgements

The authors would like to thank Picometrix for providing the LT-GaAs substrate and gratefully acknowledge the financial support from Michigan Space Grant Consortium, DARPA Young Faculty Award managed by Dr. John Albrecht, NSF CAREER Award managed by Dr. Samir El-Ghazaly, ONR Young Investigator Award managed by Dr. Paul Maki, and ARO Young Investigator Award managed by Dr. Dev Palmer.

## Author contributions

Christopher Berry designed and fabricated the plasmonic structures and photoconductive emitter prototypes. Mohammad Reza Hashemi and Mehmet Unlu assisted in the design and numerical modeling of antennas. All of the measurements are performed by Christopher Berry. Mona Jarrahi organized and supervised the project. All authors discussed the results and commented on the manuscript.

## Author Information

The authors declare no competing financial interests. Correspondence and request for materials should be addressed to Mona Jarrahi (mjarrahi@umich.edu) and Christopher W. Berry (berryc@umich.edu).

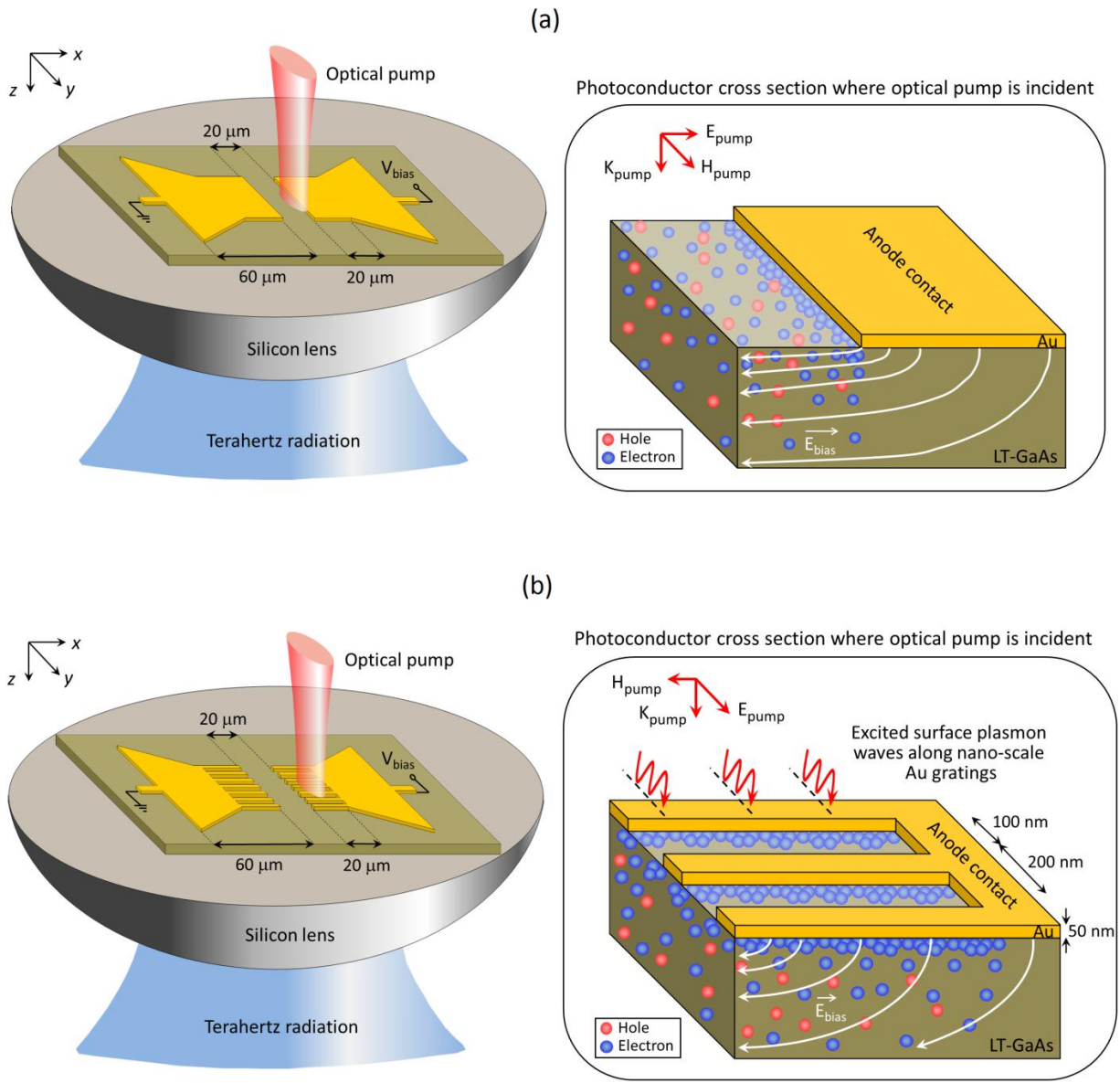

Figure 1. Schematic diagram and operation concept of photoconductive terahertz emitters. Left panels: an optical pump is incident upon an ultrafast photoconductor connected to a bowtie terahertz antenna on a LT-GaAs substrate. The device is mounted on a silicon lens to collect the terahertz radiation from back side of the substrate. Right panels: the incident optical pump generates electron-hole pairs in the substrate. Electron-hole pairs are separated under an applied bias electric field and drifted toward anode and cathode contact electrodes, respectively. Upon reaching the contact electrodes, the photocurrent drives the terahertz antenna to produce radiation. (a) A conventional photoconductive terahertz emitter. (b) A plasmonic photoconductive terahertz emitter incorporating plasmonic contact electrodes to reduce carrier transport times to contact electrodes and, thus, increase the collected photocurrent driving the terahertz antenna.

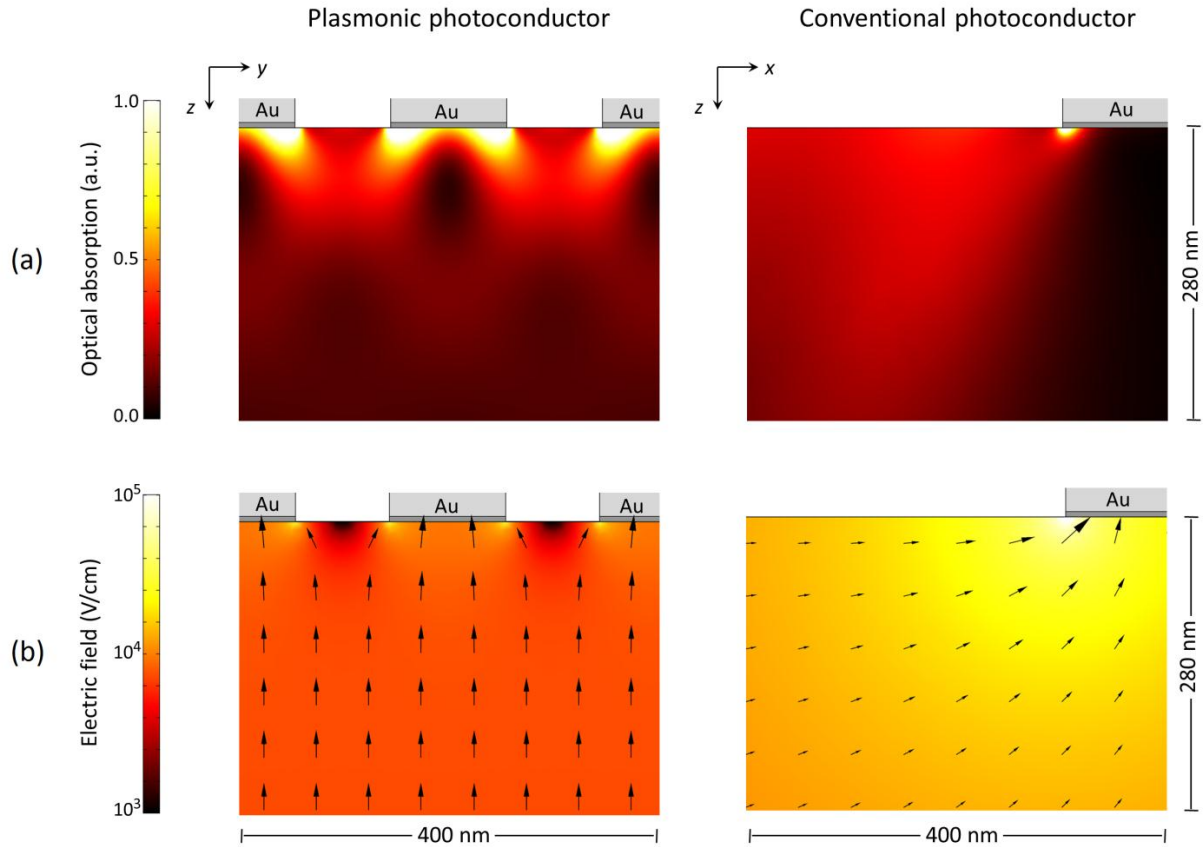

Figure 2. Finite element analysis of optical and electrical interactions in the photoconductor cross sections. Left panels: *yz*-plane plasmonic photoconductor cross sections. Right panels: *xz*-plane conventional photoconductor cross sections. (a) Color plot of optical absorption in GaAs substrate due to an incident optical plane wave ($\lambda$ = 800 nm). (b) Bias electric field in GaAs substrate. Color map corresponds to electric field magnitude, and arrows denote field direction. Applied voltages in both cases are set to not exceed electric fields of $1 \times 10^5$ V/cm. The plasmonic emitter cross section is shown for a distance 1 μm inset from the tip of the grating.

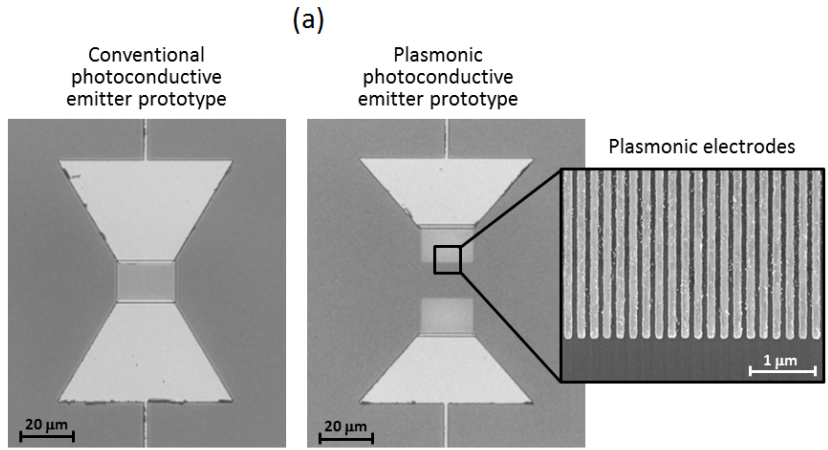

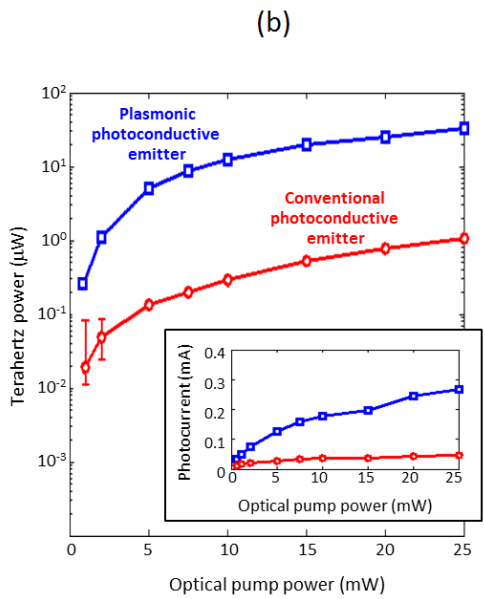
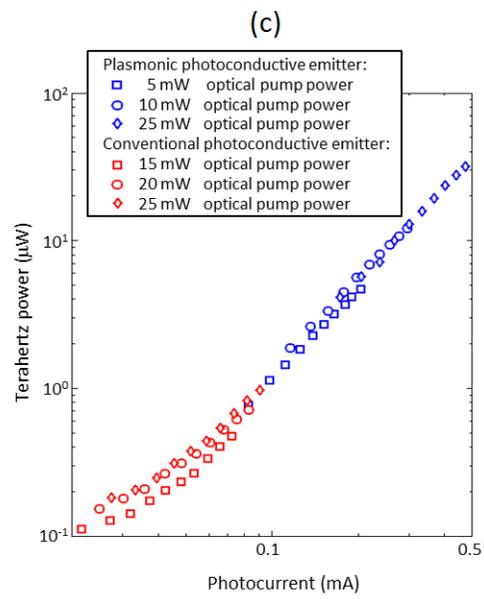
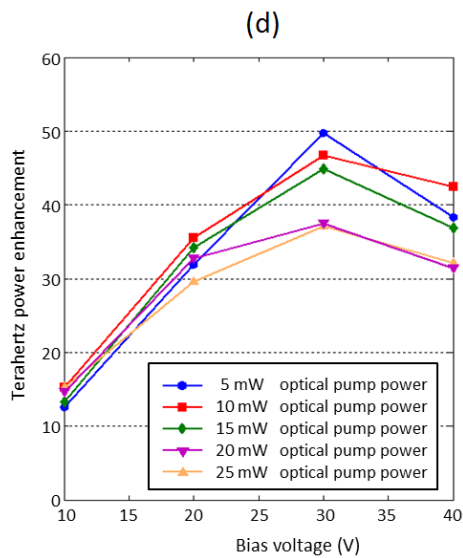
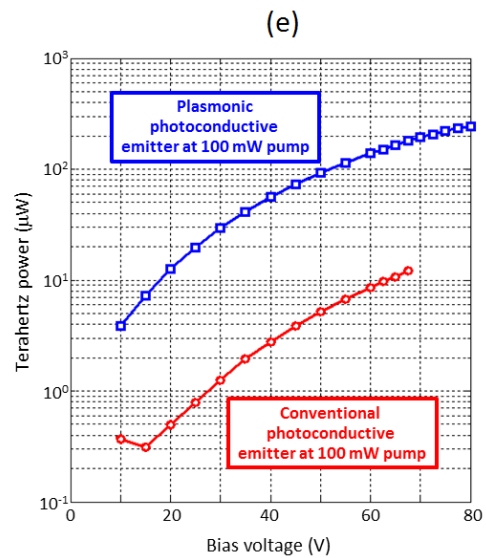

Figure 3. Comparison of conventional and plasmonic photoconductive terahertz emitters. (a) Fabricated device images. Left: microscope image of the conventional terahertz emitter. Middle: microscope image of the plasmonic terahertz emitter. Right: SEM image of the plasmonic electrodes of the plasmonic terahertz emitter. (b) Measured terahertz radiation from the plasmonic and conventional terahertz emitters, electrically biased at 40 V, under various optical pump powers. The inset curve shows the corresponding photocurrent. (c) Measured terahertz radiation versus collected photocurrent for the plasmonic and conventional terahertz emitters. The data represented in the plot includes various bias voltages (10 – 40 V) under various optical pump powers (5 – 25 mW). (d) Relative terahertz power enhancement defined as the ratio of the terahertz power emitted by the plasmonic terahertz emitter to the conventional terahertz emitter. Maximum enhancement is obtained at low optical powers before the onset of the carrier screening effect. (e) Maximum terahertz power measured from the plasmonic and conventional terahertz emitters under a 100 mW optical pump. The bias voltage of each device is increased until the point of device failure.

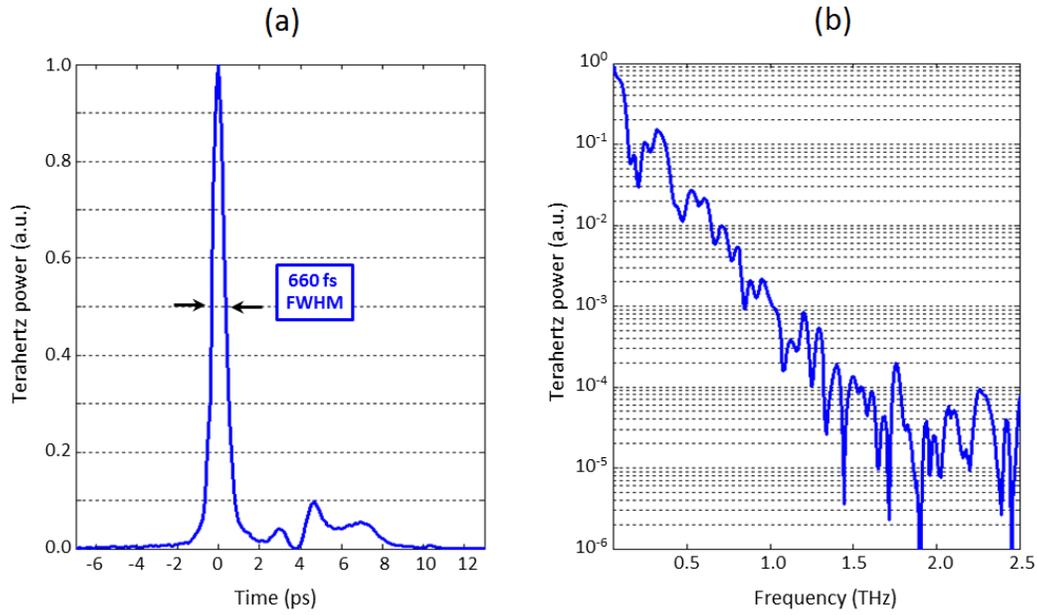

Figure 4. Radiation characteristics of the plasmonic photoconductive emitter in response to a 200 fs optical pulse from the mode-locked Ti:sapphire laser with 800 nm central wavelength and 76 MHz repetition rate. (a) Radiated power in the time domain. (b) Radiated power in the frequency domain.